\documentstyle[12pt]{article}

\input mssymb			% only necessary for integers defn
\input epsf

\advance\voffset by -2.2cm
\advance\hoffset by -2cm
\def\eq{\begin{equation}}
\def\eqe{\end{equation}}
\def\eqa{\begin{eqnarray}}
\def\eqae{\end{eqnarray}}

\def\half{\frac{1}{2}}

\def\cplex{{\mathchoice {\setbox0=\hbox{$\displaystyle\rm C$}\hbox{\hbox
to0pt{\kern0.4\wd0\vrule height0.9\ht0\hss}\box0}}
{\setbox0=\hbox{$\textstyle\rm C$}\hbox{\hbox
to0pt{\kern0.4\wd0\vrule height0.9\ht0\hss}\box0}}
{\setbox0=\hbox{$\scriptstyle\rm C$}\hbox{\hbox
to0pt{\kern0.4\wd0\vrule height0.9\ht0\hss}\box0}}
{\setbox0=\hbox{$\scriptscriptstyle\rm C$}\hbox{\hbox
to0pt{\kern0.4\wd0\vrule height0.9\ht0\hss}\box0}}}}

\def\reals{{\mathchoice {\setbox0=\hbox{$\displaystyle\rm R$}\hbox{\hbox
to0pt{\kern0.4\wd0\vrule height0.9\ht0\hss}\box0}}
{\setbox0=\hbox{$\textstyle\rm R$}\hbox{\hbox
to0pt{\kern0.4\wd0\vrule height0.9\ht0\hss}\box0}}
{\setbox0=\hbox{$\scriptstyle\rm R$}\hbox{\hbox
to0pt{\kern0.4\wd0\vrule height0.9\ht0\hss}\box0}}
{\setbox0=\hbox{$\scriptscriptstyle\rm R$}\hbox{\hbox
to0pt{\kern0.4\wd0\vrule height0.9\ht0\hss}\box0}}}}

\def\del{\partial}

\def\a{\alpha}

\def\l{\lambda}
\def\m{\mu}
\def\n{\nu}
\def\w{\omega}
\def\r{\rho}			%	\varrho
\def\s{\sigma}			%	\varsigma

\def\z{\zeta}

\def\O{\Omega}
\def\S{\Sigma}

\def\cm{{\cal M}}
\def\cv{{\cal V}}
\def\ct{{\cal T}}
\def\cz{{\cal Z}}

\def\zb{\bar{z}}
\def\cpone{\cplex P_{1}}

\def\cpn{\cplex P_{N}}
\def\cpng{\cplex P_{N_{g}}}
\def\mn{\cm_{N}}
\def\mone{M^{\prime}}

\begin{document}
\baselineskip 16pt plus 1pt minus 1pt
\hfill DAMTP 94-86
\vskip 50pt
\centerline{\bf Phase transitions in a vortex gas}
\vskip 24pt
\centerline{P.A.Shah}
\vskip 12pt
\centerline{\it Department of Applied Mathematics and Theoretical Physics}
\centerline{\it University of Cambridge}
\centerline{\it Silver Street, Cambridge CB3 9EW, U.K.}
\vskip 50pt
\centerline{\bf Abstract}
\vskip 24pt
It has been shown recently that the motion of solitons at couplings
around a critical coupling can be reduced to the dynamics of
particles (the zeros of the Higgs field) on a curved manifold with
potential. The curvature gives a velocity dependent force, and the
magnitude of the potential is proportional to the distance from
a critical coupling. In this paper we apply this approximation to
determining the equation of state of a gas of vortices in the
Abelian Higgs model. We derive a virial expansion
using certain known integrals of the metric, and the second virial
coefficient is calculated, determining the behaviour of the gas at low
densities. A formula for determining higher order coefficients is given.
At low densities and temperatures $T \gg \l$ the equation of state is of
the Van der Waals form $(P+b\frac{N^{2}}{A^{2}})(A-aN) = NT$ with
$a=4\pi$ and $b=-4.89\pi\l$ where $\l$ is a measure of the distance
from critical coupling. It is found that there is no phase transition
in a low density type-II gas, but there is a transition in the type-I
case between a condensed and gaseous state. We conclude with a discussion
of the relation of our results to vortex behaviour in superconductors.
\vfill
\vskip -12pt
\hfill September 1994
\eject

\section{Introduction}

Vortices in two space dimensions are often used to investigate
aspects of soliton behaviour as they exhibit many of the features
of more complex solitons, and the low dimensional Abelian nature
of the theory makes it simpler to investigate \cite{taubes}.
They are the products of spontaneous symmetry breaking in the
Abelian Higgs model, and have (zero temperature) masses of the
order of the symmetry breaking scale, and so with their extended size,
can be regarded as ``almost classical'' objects.
\par
Apart from the mathematical interest, the model also has a physical
significance. The Abelian Higgs model is also known as the
Landau-Ginzburg model in the theory of superconductivity, and the
motion of vortices has a dramatic effect on the
properties of the material. Although a three
dimensional problem, there are indications that the interactions
of the flux tubes are well modelled by the two dimensional
planar interaction of vortices
\cite{nelson}. Although thin film superconductors are essentially
two dimensional, the interaction of the magnetic field external to
the superconductor leads to an extra static force proportional to the
logarithm of the distance between the vortices; we do not consider
this additional effect in this paper. We also note that vortices may
occur as products of spontaneous symmetry breaking in the early
universe. Cosmic strings (as they are known) are potentially relevant in
theories of galaxy formation in the early universe through their
gravitational effect on spacetime.
\par
At a critical value of a coupling constant in the Lagrangian of many
theories containing soliton solutions (including the Abelian Higgs
model), static multi-soliton solutions exist
with the solitons residing at arbitrary
spatial positions and internal coordinates; the {\it moduli}.
The manifold formed by these solutions is termed the {\it moduli space}.
Away from this coupling, soliton states can still be identified, but
the solitons typically exert forces on each other, and static
solutions are isolated (for example, the famous Abrikosov flux lattice
\cite{abrikosov}).
\par
A difficult task is to determine the nature of soliton dynamics,
especially for theories where the solutions of the field
equations cannot be written in closed form, but must be computed
numerically. One approach that has attracted study is to put the field
equations on a lattice, and to simulate the field theory on a
computer. This approach has been fruitful in studying the interactions
of vortices and strings, but is however limited to
describing the behaviour of small numbers of solitons (usually just
two) at definite initial velocities, coordinates and coupling,
and cannot provide a general description of soliton scattering or
their statistical behaviour.
\par
In a recent paper by this author \cite{ncv}, it was shown how the
field dynamics of a configuration of vortices at close to critical
coupling is well approximated by a finite dimensional Lagrangian
particle evolution in the moduli space, where the coordinate in the
moduli space marks the positions of the centres of a vortices
(although the Abelian Higgs model was used, the method could be
applied to other theories). This generalises an earlier result
due to Manton \cite{geoapprox} for vortices at critical
coupling. The motion is determined by a non-trivial metric,
and a potential which vanishes at critical coupling. The metric
and potential were derived by an analogy with
the paths of steepest descent from unstable to
stable static solutions, in the ``true'' configuration space
of the theory (that is, the space of all gauge orbits of suitably
smooth field configurations).
\par
The curvature of the metric gives non-trivial scattering
even when the static forces vanish between vortices (and,
as we shall see later, a modification in physical properties
of a vortex gas); the motion is geodesic in this case. At non-critical
coupling, the potential leads to a forcing term away from
geodesic motion. We expect this situation to be generic for all
couplings, although in the extreme type-II region the potential may be
modified by terms extra to those we consider here. It is also
important to note that our model {\it does not} require the use of
limits where the vortex core size is taken to zero
(the London limit), or where the magnetic field is close to
its maximum allowed value (the Landau limit).
\par
The Abelian Higgs model has spontaneously broken $U(1)$ gauge
symmetry, and can be defined on any orientable Riemann surface. We
are considering either manifolds with no boundary, or a boundary
at infinity, so we are able to discuss the dynamics of type-I
vortices.  We take spacetime locally to have the metric
\eq
ds^{2} = dx_{0}^{2} - \O^{2}(x_{1},x_{2})(dx_{1}^{2} + dx_{2}^{2})
\eqe
where $x_{1}$ and $x_{2}$ are suitably chosen local coordinates (later
on, we will find it more convenient to use the complex coordinate
defined by $z=x_{1} + i x_{2}$). We will define $\O^{2}$ in the next
section when we specify the thermodynamic model. The field action is
\eq
\label{eq:action1}
S = \int \O^{2} d^{3}x [ -\frac{1}{4}f_{\m\n}f^{\m\n} + \half D^{\m}\phi
(D_{\m}\phi)^{*} + \a (|\phi|^{2} - c_{0}^{2})^{2}]
\eqe
where $\phi$ is a complex scalar valued Higgs field (in the
Landau-Ginzburg model of superconductivity, $\phi$ is termed the
{\it order parameter}). $D_{\m}\phi =
(\del_{\m} -ieA_{\m})\phi$ where $A_{\m}$ is the real valued gauge
potential, and $e$ the gauge coupling. We are considering a classical
field theory here, but in a (semiclassical) finite temperature
effective quantum field theory the constants $c_0$, $e$ and $\l$
are all dependent on temperature. We shall not discuss this
dependence here, but take the temperature to be constant.
The action (\ref{eq:action1}) can be rescaled using the
dimensionless quantities
\eq
x^{\m} = \frac{1}{ec_{0}}\tilde{x}^{\m}, \;\;\; \phi =
c_{0}\tilde{\phi}, \;\;\; A_{\m} = c_{0}\tilde{A_{\m}}, \;\;\;
S = \frac{c_{0}}{e} \tilde{S}
\eqe
to read
\eq
\label{eq:action}
\tilde{S} = \int \O^{2}d^{3}\tilde{x} [ -\frac{1}{4}\tilde{f}_{\m\n}
\tilde{f}^{\m\n} +
\half \tilde{D}^{\m}\tilde{\phi} (\tilde{D}_{\m}\tilde{\phi})^{*} +
(\frac{1}{8} +\l)(|\tilde{\phi}|^{2} - 1)^{2}]
\eqe
where $\frac{1}{8} +\l = \frac{\a}{e^{2}}$ (so $\l > -\frac{1}{8}$ for
stability of the vacuum) and $\tilde{D_{\m}}\tilde{\phi} = (\tilde{\del_{\m}} -
i\tilde{A_{\m}})\tilde{\phi}$. $\l=0$ is the critical coupling for
this model. In these units the mass of the gauge
particle $m_{``photon''} = 1$, and the mass of the scalar field is
$m_{Higgs} = \sqrt{1+8\l}$. For a superconductor, the units of
physical quantities can be easily recontructed from the known
properties of the material.
\par
$\phi$ and $A_{i}$ are not smooth
functions on the Riemann surface, but rather a section and connection on a
non-trivial bundle over it with first Chern number $N$.
Thus any finite energy configuration has a topologically quantised
magnetic flux
\eq
\int d^{2}x f_{12} = 2\pi N
\eqe
and this magnetic flux is localised in the neighbourhoods of the
$N$ zeros of $\phi$, which we identify as the positions of the vortices.
\par
In \cite{ncv}, it was shown that the dynamics of a configuration
$(\phi,A_{i})$ (we use the gauge fixing condition $A_{0} =0$ which
is sufficient for our purposes) is well approximated for small $\l$ by
\eqa
\label{eq:themodel}
S & = & \int dx_{0} ( \ct - \cv) \\
\ct & =  & \sum_{i,j=1}^{N} \half m g_{ij}(Q)
\frac{dz^{i}}{dt} \frac{d\zb^{j}}{dt} \nonumber \\
\cv & = & N\pi + U_{N}(Q) \nonumber
\eqae
where $Q$ is a collective coordinate on the moduli space encapsulating
the (complex) coordinates $\{z_{1}, \ldots ,z_{N}\}$ of the zeros of the
Higgs field $\phi$. The metric $g_{ij}(Q)$ is the induced metric on
the moduli space {\it at critical coupling}, which can
be expressed in terms of derivatives of the critically coupled
field configuration $\phi^{(0)}(Q)$ which has the same set of zeros
$\{z_{1}, \ldots ,z_{N}\}$. This metric has been shown to be
K\"ahler \cite{samols}, a geometrical property which will be essential
to our determination of the thermodynamics of vortices.
$m$ is the total energy of the $N=1$ solution at near-critical
coupling, which may be identified as the inertial mass of an isolated
vortex. This energy may be computed numerically to
be
\eq
m = \pi + 5.21\l +O(\l^{2})
\eqe
The potential $U_{N}(Q)$ can be expressed as the
integral
\eq
\label{eq:vortpot}
U_{N}(Q) = \l \int (|\phi_{0}(Q)|^{2} - 1)^{2} \O^{2}d^{2}x \; +O(\l^{2})
\eqe
We retain only the lowest order terms in $\l$. Note that this is
intrinsically an $N$-body potential, rather than a sum of pair
interactions (we shall see later that this would not model the vortex
interactions satisfactorily). As the separation of the vortices
increases, $U_{N} \rightarrow N(m-\pi)$. We normalise the
potential by subtracting off this value.
$U_{N}(Q) \rightarrow U_{N}(Q) - N(m-\pi)$. This model is supported
in part by the recent rigourous results of Stuart \cite{stuart},
and has been tested against lattice computations of two vortex
scattering with good agreement for velocities up to around $0.4c$
and couplings of half or twice the critical value \cite{ncv}.
\par
Unfortunately, neither the metric nor the potential can be calculated
in closed form for $N \ge 2$ vortices (except in the special cases
where the number density of vortices approaches its maximum allowed
value (the Landau limit) \cite{hidensity}, and the case of vortices
on a hyperbolic surface with a particular negative curvature
\cite{strachan}). It can be calculated numerically for
two vortices, but even this one dimensional
problem is time consuming to solve. However, in the next three
sections we describe how a choice of formulation of the
thermodynamical model enables symmetry properties of the system to be
exploited, and a virial expansion for the system calculated in which
at least the second virial coefficient may be calculated from known
data. Using this result, we analyse the physics of a low density gas
of vortices, determining that it can undergo a discontinuous phase transition
when the vortices are attracting. We also comment on estimates for the
third and higher virial coefficients, and how the predicted
behaviour of the vortex gas may be modified by their inclusion.
We conclude with a summary of our results and a discussion of recent
experiments with high temperature superconductors.

\section{The thermodynamical model}

The partition function for a gas of $N$ critically coupled vortices on the
sphere $S^{2}$ (or conformally compactified complex plane
$\cplex \cup \{ \infty \}$) with area $A$ has been calculated in
\cite{smv}. The sphere was used as its special symmetry simplified the
calculation, but in the thermodynamic limit $N \rightarrow \infty$ at
fixed number density $n=N/A$ the physical properties of the gas of
vortices is expected to be insensitive to the topology of the surface
they are defined upon. Thus the thermodynamics of vortices on a sphere
(where the thermodynamic limit can be better controlled)
models the thermodynamics on the plane. This was supported in
\cite{tvp}, where the partition function for a gas of critically
coupled vortices in a section of the plane with periodic boundary
conditions and finite area (i.e. a torus) was calculated and the
physical properties were found to be equal to the results for the
sphere with $O(1/N)$ corrections.
\par
In our derivation of the equation of state for a near-critical
vortex gas, we will also use the sphere in the similar expectation
that it will also model the thermodynamics of vortices on the plane.
The sphere is chosen as its special symmetry enables the calculation
of some integrals in the partition function. To properly handle the
effect of the potential between the vortices, we will derive a virial
expansion for the equation of state, and this requires the use of the
grand canonical partition function. In this system, the particles are
in contact with a heat bath of temperature $T$ (we assume the other
field modes are thermalised also) and an infinite source/sink of particles.
\par
We take the number of vortices on the sphere to be $N_{g}$, which is
fixed topologically. To define the grand canonical partition function,
we work in the limit $N_{g} \rightarrow \infty$, and define an open
region $M$ of the sphere as the entire sphere, but with a point
deleted: $M = S^{2} - \{ \infty \} \simeq \cplex$. $M$ has the
same area as $S^{2}$, which we take to be $A$. In this case,
$\O(z,\zb)^{2} = A/ \pi(1+ z \zb)^{2}$ where $z$ is the complex
coordinate in the plane.
\par
An important aspect of the geometric structure of the moduli
space for $N_{g}$ vortices on the sphere is due to the
vortices being classically indistinguishable particles. We may obtain
a coordinatisation of the moduli space by considering the
polynomial equation
\eq
a_{N_{g}}z^{N_{g}} + a_{N_{g}-1}z^{N_{g}-1} + \ldots + a_{1}z + a_{0}
= 0
\eqe
The roots of this polynomial are an unordered set $\{ z_{1}, \ldots
,z_{N_{g}} \}$. If $a_{N_{g}} = 0$ then this is interpreted as one of
the roots being infinity, and so on. Multiplying each $a_{i}$ by the
same complex number $c \neq 0$ does not change the roots of the
polynomial, so the set $Q = \{a_{N_{g}}, \ldots, a_{0} \}$ is seen to
provide homogenous coordinates for the moduli space which is
the complex projective space $\cpng$.
\par
In order to properly control the thermodynamic limit, we make use of
two submanifolds of this. Let
$\mn$ be the manifold obtained by letting $N \le N_{g}$ of the
vortices vary around $\cplex$ whilst keeping the other $N_{g} - N$
fixed at $\infty$ i.e. $\mn \simeq \cplex^{N}/\S_{N}$ where $\S_{N}$
is the permutation group of $N$ points. This can be defined in
terms of the homogenous coordinates as the submanifold obtained
by the restriction $a_{N_{g}} = \ldots = a_{N+1} = 0$ , $a_{N} \neq
0$. A submanifold $\mone$ of this, isomorphic to the complex plane $M$,
is obtained in the following way: choose an arbitrary
labelling for the positions of the $N$ vortices, $\{ z_{1}, \ldots ,
z_{N}\}$ and let $\mone$ be the surface obtained by allowing one of the
labelled vortices $z_{i}$ to vary over the complex plane, keeping the others
fixed. The dependence of the homogenous coordinates on $z_{i}$ is
linear, so $\mone$ is a complex projective line $\cpone$ with one
point removed, in $\cpng$. $\mone \simeq \cplex$ is the fundamental
range of integration for our coordinates, and the labelling is
introduced so that the potential may be defined as (permutation
invariant) functions of the vortex positions, rather than functions
on $\cpn$. This avoids a technical difficulty of
lifting potentials defined on $\cpn$ to $\cplex P_{N+j}$.
\par
It is a remarkable corollary of the  K\"ahler property of $\cpng$ that
the area of $\mone$ may be calculated even without an explicit
knowledge of the area measure on it. The area measure is derived from
the K\"ahler form $\w$ defined as
\eq
\w = \frac{1}{2i} g_{ij} dz^{i} \wedge d\zb^{j}
\eqe
where $g_{ij}$ is the metric on $\cpng$ given by the kinetic energy of
a drift motion in the moduli space in equation (\ref{eq:themodel}).
The restriction of $\w$ to a surface in $\cpng$ is the area form on
that surface. The K\"ahler form is closed, the area $[\w]$ of all
$\cpone$ lines are equal, and consideration of the special motion of
$N_{g}$ coincident vortices \cite{smv} allows this area to be deduced
as
\eq
\label{eq:area}
[\w] = A - 4\pi N_{g}
\eqe
where $A > 4\pi N_{g}$ is required a priori \footnote{This is a
theoretical constraint, but is equivalent to the requirement that we
are below the upper critical magnetic field density.}
for the existence of vortex solutions \cite{bradlow}. The subtraction
of the term proportional to the number of vortices $N_{g}$
is a consequence of the curvature of the metric. We note that
\eq
\mbox{Area($\mone$)} = [\w]
\eqe
This is our motivation for choosing $\mone$ as a $\cpone$ line minus
a point. The dependence of the area of $\mone$ on the
{\it total} number of vortices on the sphere, including the ones
outside $\mone$ is perhaps surprising, but it is a consequence of
our definition of $\mone$ as being the whole sphere minus a point.
However, it will not affect the thermodynamic limit of our results.
\par
We are now in a position to define the grand canonical partition
function for our problem. It is
\eqa
\label{eq:gcpf1}
\cz & = & \sum_{N=0}^{\infty} \frac{1}{N!}\z^{N} Z_{N} \\
Z_{N} & = & \int g^{\half} dz_{1} \wedge d\zb_{1} \ldots \int g^{\half}
dz_{N} \wedge d\zb_{N} \; \exp(-U_{N}(z_{1} \ldots z_{N})/T) \nonumber
\eqae
where
\eq
\z = \frac{2\pi m T}{h^{2}} e^{\frac{\m}{T}}
\eqe
is the {\it activity} of the gas obtained by integrating out the
momentum coordinates in the Hamiltonian ($\m$ is a chemical
potential and $h$ is Planck's constant) defined (in collective
coordinates) by
\eq
H= \frac{1}{2m} g_{ij}(Q)P^{i}P^{j} + U_{N}(Q)
\eqe
where $P^{i} = m\dot{Q}^{i}$ is the conjugate momentum coordinate.
The range of integration of each $z_{i}$ is $\mone \simeq
\cplex$, and $g^{\half}$ is the area measure on each $\mone$.
$U_{N}$ is the potential function for $N$ vortices. All functions are
symmetric functions of the $z_{i}$, and the factor $1/N!$ must be
inserted to allow for the indistinguishable nature of the vortices
(this factor would arise naturally without the need for insertion ``by
hand'' if we were using $\cplex^{N}/\S_{N}$ as our range of
integration, rather than $\cplex^{N}$).

\section{The virial expansion of equation of state}

The virial expansion is obtained by a resummation, or regrouping, of
terms in the partition function into {\it cluster functions} which
encapsulate the interactions of small numbers, or clusters of
vortices. The derivation of the virial coefficients that we adopt here
mirrors the standard textbook derivation \cite{mayer}, but with two important
differences: 1) the potential function is not well approximated by a
sum of pair terms, so we must in principle keep higher point terms
in any expansion for it; 2) the surfaces $\mone$ integrated over are not flat,
and their metrics are not known (although as indicated above we can
determine their area).
\par
Define
\eq
W_{N} = \exp(-\frac{U_{N}}{T})
\eqe
and expand $U_{N}$ as a sum
\eq
U_{N} = \sum_{N \ge i > j \ge 1} u_{2}(z_{i},z_{j}) +
        \sum_{N \ge i > j > k \ge 1} u_{3}(z_{i},z_{j},z_{k})
        + \ldots + u_{N}(z_{1}, \ldots , z_{N})
\eqe
where
\eqa
u_{2}(z_{i},z_{j}) & = & U_{2}(z_{i},z_{j}) \\
u_{3}(z_{i},z_{j},z_{k}) & = & U_{3}(z_{i},z_{j},z_{k}) -
U_{2}(z_{i},z_{j}) - U_{2}(z_{j},z_{k}) - U_{2}(z_{i},z_{k}) \nonumber
\eqae
are 2-body and 3-body potentials and so on. This terminates correctly
for any $N$, and has the advantage that each $u_{m}$ is only non-zero
when at least $m$ vortices are close together.
It can be shown, for example, that
$U_{3}(z_{i},z_{j},z_{k}) \rightarrow U_{2}(z_{i},z_{j}) + O(\exp
(-|z_{k}-z_{i}|), \exp(-|z_{k}-z_{j}|))$  when $z_{i}$ and $z_{j}$ are
close and $z_{k}$ is separated from the other two. At this stage, we
need to keep all terms in the series as, for example $\sum_{i>j}
u_{2}(z_{i},z_{j})$ does not approximate $U_{N}$ well when
more than two vortices are close. Defining
\eq
f_{i_{1} \ldots i_{m}}(z_{i_{1}}, \ldots ,z_{i_{m}}) = e^{-u_{m}
(z_{i_{1}}, \ldots ,z_{i_{m}})/T} - 1
\eqe
we have
\eqa
\label{eq:wexpand}
W_{N} & = & [\prod_{i>j}^{N} (1+f_{ij})]\;[\prod_{i>j>k}^{N}
(1+f_{ijk})] \ldots \\
 & = & [1 + \sum_{i>j}^{N}f_{ij} + \sum_{i>j,k>l}^{N}f_{ij}f_{kl} +
\ldots ]\;[1 + \sum_{i>j>k}^{N}f_{ijk} +
\sum_{i>j>k,l>m>n}^{N}f_{ijk}f_{lmn} + \ldots ] \ldots \nonumber
\eqae
We now rephrase the grand canonical partition function as the
exponential of a sum of cluster functions $V_{l}$
\eq
\label{eq:pfclus}
\cz = \exp ( \,\sum_{l=1}^{\infty} \frac{\z^{l}}{l!} \int g^{\half}
dz_{1} \wedge d\zb_{1} \ldots \int g^{\half} dz_{l} \wedge d\zb_{l}
\; V_{l} (z_{1}, \ldots , z_{l}))
\eqe
By expanding (\ref{eq:pfclus}) and equating terms in $\z^{N}$ with
(\ref{eq:gcpf1}) and using the permutation symmetry of the functions
in terms of their arguments, we find that for the first few terms
\eqa
V_{1}(z_{1}) & = & W_{1}(z_{1}) \\
V_{2}(z_{1},z_{2}) & = & W_{2}(z_{1},z_{2})- W_{1}(z_{1})W_{1}(z_{2})
\nonumber \\
V_{3}(z_{1},z_{2},z_{3}) & = & W_{3}(z_{1},z_{2},z_{3}) -
W_{1}(z_{1})W_{2}(z_{2},z_{3}) - W_{1}(z_{2})W_{2}(z_{3},z_{1})-
\nonumber \\
& & W_{1}(z_{3})W_{2}(z_{1},z_{2}) + 2
W_{1}(z_{1})W_{1}(z_{2})W_{1}(z_{3}) \nonumber \\
\eqae
where we have used the permutation invariance of the various functions
and integrated over free coordinates to make these identifications.
Using the expansion for $W_{N}$ given in (\ref{eq:wexpand}) this
can be written
\eqa
V_{1} & = & 1 \\
V_{2} & = & f_{12} \nonumber \\
V_{3} & = & f_{12}f_{23} +f_{13}f_{12} +f_{13}f_{23}
+f_{12}f_{23}f_{13} + \nonumber \\
& & f_{123}(1+f_{12}+f_{23}+f_{13}+f_{12}f_{23} +f_{13}f_{12}
+f_{13}f_{23} +f_{12}f_{23}f_{13}) \nonumber
\eqae
The functions $W_{N}$ and $V_{N}$ can be denoted graphically. The main
difference from the usual diagrams is the presence of terms in
addition to the sum of pair terms $u_{2}$. These can be denoted as
different ``colours''; for example the three point terms $f_{ijk}$
 can be denoted as
``red'' triangles joining combinations of multiples of 3 points, four
point terms $f_{ijkl}$ could be denoted as ``green'' combinations of
multiples of four points, and so on. Thus, the usual interpretation of
the cluster functions $V_{l}$ as a sum of graphs consisting of $l$
points that are connected is replaced by a sum over graphs
consisting of $l$ points {\it connected in at least one colour $\le l$
} (where ``black'' is given the value $2$, ``red'' is $3$, ``green''
is $4$ and so on). Colour screens and printers are unfortunately rare,
so we have denoted ``red'' here by dotted lines (``green and higher
point terms are left as an exercise for the reader''). The graphical
expression of the first three $W_{l}$ terms is:
\eqa
W_{1} & = & \mbox{
\begin{picture}(4,4)(18,828)
\thinlines
\put( 20,830){\circle*{4}}
\end{picture}
}
\nonumber \\
W_{2} & = & \mbox{
\begin{picture}(84,15)(18,827)
\thinlines
\put( 40,830){\circle*{4}}
\put( 79,830){\circle*{4}}
\put( 20,830){\circle*{4}}
\put(100,830){\circle*{4}}
\put( 57,827){\makebox(0,0)[lb]{\raisebox{0pt}[0pt][0pt]{\twlrm +}}}
\thicklines
\put( 79,830){\line( 1, 0){ 21}}
\end{picture}
}
\nonumber \\
W_{3} & = & \mbox{
\begin{picture}(352,20)(55,807)
\thicklines
\put( 68,800){\circle*{4}}
\put(140,820){\circle*{4}}
\put(153,800){\circle*{4}}
\put( 80,820){\circle*{4}}
\put(187,800){\circle*{4}}
\put(213,800){\circle*{4}}
\put(260,820){\circle*{4}}
\put(247,800){\circle*{4}}
\put(273,800){\circle*{4}}
\put(320,820){\circle*{4}}
\put(333,800){\circle*{4}}
\put(380,820){\circle*{4}}
\put(367,800){\circle*{4}}
\put(393,800){\circle*{4}}
\put(307,800){\circle*{4}}
\put(200,820){\circle*{4}}
\put(127,800){\circle*{4}}
\put( 93,800){\circle*{4}}
\put(260,820){\line(-2,-3){ 13.231}}
\put(247,800){\line( 1, 0){ 26}}
\put(273,800){\line(-2, 3){ 13.231}}
\put(200,819){\line(-2,-3){ 12.769}}
\put(187,800){\line( 1, 0){ 26}}
\put(127,800){\line( 1, 0){ 26}}
\multiput(380,820)(-4.41027,-6.61540){4}{\makebox(0.4444,0.6667){\tenrm .}}
\multiput(367,800)(8.66667,0.00000){4}{\makebox(0.4444,0.6667){\tenrm .}}
\multiput(393,800)(-4.41027,6.61540){4}{\makebox(0.4444,0.6667){\tenrm .}}
\put( 55,806){\makebox(0,0)[lb]{\raisebox{0pt}[0pt][0pt]{\twlrm (}}}
\put(407,806){\makebox(0,0)[lb]{\raisebox{0pt}[0pt][0pt]{\twlrm )}}}
\put(347,806){\makebox(0,0)[lb]{\raisebox{0pt}[0pt][0pt]{\twlrm + }}}
\put(286,806){\makebox(0,0)[lb]{\raisebox{0pt}[0pt][0pt]{\twlrm ) ( }}}
\put(232,806){\makebox(0,0)[lb]{\raisebox{0pt}[0pt][0pt]{\twlrm +}}}
\put(167,806){\makebox(0,0)[lb]{\raisebox{0pt}[0pt][0pt]{\twlrm +3}}}
\put(107,807){\makebox(0,0)[lb]{\raisebox{0pt}[0pt][0pt]{\twlrm +3}}}
\end{picture}
}
\nonumber
\eqae
which implies the first few $V_{l}$ terms are
\eqa
V_{1} & = & \mbox{
\begin{picture}(4,4)(18,828)
\thinlines
\put( 20,830){\circle*{4}}
\end{picture}
}
\nonumber \\
V_{2} & = & \mbox{
\begin{picture}(24,4)(18,828)
\thinlines
\put( 20,830){\circle*{4}}
\put( 40,830){\circle*{4}}
\thicklines
\put( 21,830){\line( 1, 0){ 20}}
\end{picture}
}
\nonumber \\
V_{3} & = & \mbox{
\begin{picture}(407,20)(55,807)
\thicklines
\put( 68,800){\circle*{4}}
\put( 93,800){\circle*{4}}
\put(140,820){\circle*{4}}
\put(127,800){\circle*{4}}
\put(153,800){\circle*{4}}
\put( 80,820){\circle*{4}}
\put(187,800){\circle*{4}}
\put(213,800){\circle*{4}}
\put(260,820){\circle*{4}}
\put(247,800){\circle*{4}}
\put(273,800){\circle*{4}}
\put(320,820){\circle*{4}}
\put(307,800){\circle*{4}}
\put(333,800){\circle*{4}}
\put(380,820){\circle*{4}}
\put(367,800){\circle*{4}}
\put(393,800){\circle*{4}}
\put(440,820){\circle*{4}}
\put(427,800){\circle*{4}}
\put(453,800){\circle*{4}}
\put(200,820){\circle*{4}}
\put( 80,820){\line(-3,-5){ 12}}
\put( 68,800){\line( 1, 0){ 25}}
\put(140,820){\line(-3,-5){ 12}}
\put(128,800){\line( 1, 0){ 25}}
\put(153,800){\line(-2, 3){ 13.231}}
\put(308,800){\line( 1, 0){ 25}}
\put(380,820){\line(-2,-3){ 13.231}}
\put(367,800){\line( 1, 0){ 26}}
\put(440,819){\line(-2,-3){ 12.769}}
\put(427,800){\line( 1, 0){ 26}}
\put(454,800){\line(-2, 3){ 13.539}}
\multiput(200,820)(-4.41027,-6.61540){4}{\makebox(0.4444,0.6667){\tenrm .}}
\multiput(187,800)(8.66667,0.00000){4}{\makebox(0.4444,0.6667){\tenrm .}}
\multiput(213,800)(-4.41027,6.61540){4}{\makebox(0.4444,0.6667){\tenrm .}}
\put(107,807){\makebox(0,0)[lb]{\raisebox{0pt}[0pt][0pt]{\twlrm +}}}
\put(407,806){\makebox(0,0)[lb]{\raisebox{0pt}[0pt][0pt]{\twlrm +}}}
\put(462,806){\makebox(0,0)[lb]{\raisebox{0pt}[0pt][0pt]{\twlrm )}}}
\put(347,806){\makebox(0,0)[lb]{\raisebox{0pt}[0pt][0pt]{\twlrm + 3}}}
\put(286,806){\makebox(0,0)[lb]{\raisebox{0pt}[0pt][0pt]{\twlrm + 3}}}
\put(232,806){\makebox(0,0)[lb]{\raisebox{0pt}[0pt][0pt]{\twlrm (}}}
\put(167,806){\makebox(0,0)[lb]{\raisebox{0pt}[0pt][0pt]{\twlrm +}}}
\put( 55,806){\makebox(0,0)[lb]{\raisebox{0pt}[0pt][0pt]{\twlrm 3}}}
\end{picture}
}
\nonumber
\eqae
Because of base point invariance (the choice of origin of
coordinates), each cluster integral is proportional to $[\w]$. Writing
\eq
b_{l} = \frac{1}{l! [\w]} \int g^{\half} dz_{1} \wedge d\zb_{1}
\ldots \int g^{\half} dz_{l} \wedge d\zb_{l} \; V_{l}(z_{1},\ldots ,z_{l})
\eqe
then $b_{l}$ depends at most on the {\it effective density} $\r =
\frac{<N>}{[\w]} $. We are now in a position to write down a virial
expansion for the equation of state in terms of the effective density.
The grand canonical potential $\O = -T \ln \cz$ is
\eq
\label{eq:gcp}
\O = -T [\w] \sum_{l=1}^{\infty} \z^{l} b_{l}
\eqe
The expectation number of vortices is $<N>= -\frac{\del \O}{\del \m}$ and
the pressure $P = - \frac{\del \O}{\del A}$. These have series expansions
\eqa
\label{eq:avn}
<N> = [\w] \sum_{l=1}^{\infty} l b_{l} \z^{l} \\
\label{eq:pressz}
P = T \sum_{l=1}^{\infty} b_{l} \z^{l}
\eqae
The virial expansion of the equation of state is
\eq
\label{eq:virialexp}
\frac{P[\w]}{<N> T} = \sum_{l=1}^{\infty} B_{l} \r^{l-1}
\eqe
Substituting (\ref{eq:avn}) and (\ref{eq:pressz}) into
(\ref{eq:virialexp}) and equating terms in $\z^{j}$ we obtain for the
first few virial coefficients $B_{l}$
\eqa
B_{1} & = & b_{1} = 1 \\
B_{2} & = & -b_{2} \nonumber \\
B_{3} & = & 4b_{2}^{2} - 2b_{3} \nonumber \\
B_{4} & = & -20b_{2}^{3} + 18b_{2}b_{3} - 3b_{4} \nonumber
\eqae
Because of the K\"ahler property of the moduli space, these formulae
are identitical to those obtained in flat space. The effect of the
curvature is coded into the integrals for the $b_{l}$. In the
thermodynamic limit we may identify $<N>$ and $N_{g}$, and write
it as $N$. The equation of state is then
\eq
\label{eq:pressseries}
P = T \sum_{l=1}^{\infty} B_{l}\r^{l}
\eqe
The coefficients $B_{l}$ for $l \ge 2$ can be expressed as
a series in $\l/T$ without a constant term, so as $\l \rightarrow 0$,
the equation of state smoothly approaches the critically coupled
equation calculated in \cite{smv} and \cite{tvp}.
\par
The virial coefficients can be denoted graphically in the same manner
as the cluster functions. $B_{l}$ can be written as
\eq
\label{eq:vcoff}
B_{l} = -\frac{1}{[\w]l(l-2)!} \int g^{\half} dz_{1} \wedge d\zb_{1}
\ldots \int g^{\half} dz_{l} \wedge d\zb_{l} \; X_{l}(z_{1},\ldots,z_{l})
\eqe
where $X_{l}$ is a ``modified'' {\it star function}, that is a sum
over all terms corresponding to an $l$-point graph that is connected
in at least one colour $\le l$ if we remove any point of the graph.
Thus
\eqa
\label{eq:x}
X_{2} & = & f_{12} \\
& = & \mbox{
\begin{picture}(24,4)(18,828)
\thinlines
\put( 20,830){\circle*{4}}
\put( 40,830){\circle*{4}}
\thicklines
\put( 21,830){\line( 1, 0){ 20}}
\end{picture}
} \nonumber \\
X_{3} & = & f_{12}f_{13}f_{23} + f_{123}(1 +f_{12}+f_{23}+f_{13}
+f_{12}f_{23}+f_{13}f_{12}+f_{13}f_{23} +f_{12}f_{13}f_{23}) \nonumber
\\
& = & \mbox{
\begin{picture}(337,20)(125,807)
\thicklines
\put(140,820){\circle*{4}}
\put(127,800){\circle*{4}}
\put(153,800){\circle*{4}}
\put(187,800){\circle*{4}}
\put(213,800){\circle*{4}}
\put(260,820){\circle*{4}}
\put(247,800){\circle*{4}}
\put(273,800){\circle*{4}}
\put(320,820){\circle*{4}}
\put(307,800){\circle*{4}}
\put(333,800){\circle*{4}}
\put(380,820){\circle*{4}}
\put(367,800){\circle*{4}}
\put(393,800){\circle*{4}}
\put(440,820){\circle*{4}}
\put(427,800){\circle*{4}}
\put(453,800){\circle*{4}}
\put(200,820){\circle*{4}}
\put(140,820){\line(-3,-5){ 12}}
\put(128,800){\line( 1, 0){ 25}}
\put(153,800){\line(-2, 3){ 13.231}}
\put(308,800){\line( 1, 0){ 25}}
\put(380,820){\line(-2,-3){ 13.231}}
\put(367,800){\line( 1, 0){ 26}}
\put(440,819){\line(-2,-3){ 12.769}}
\put(427,800){\line( 1, 0){ 26}}
\put(454,800){\line(-2, 3){ 13.539}}
\multiput(200,820)(-4.41027,-6.61540){4}{\makebox(0.4444,0.6667){\tenrm .}}
\multiput(187,800)(8.66667,0.00000){4}{\makebox(0.4444,0.6667){\tenrm .}}
\multiput(213,800)(-4.41027,6.61540){4}{\makebox(0.4444,0.6667){\tenrm .}}
\put(407,806){\makebox(0,0)[lb]{\raisebox{0pt}[0pt][0pt]{\twlrm +}}}
\put(462,806){\makebox(0,0)[lb]{\raisebox{0pt}[0pt][0pt]{\twlrm )}}}
\put(347,806){\makebox(0,0)[lb]{\raisebox{0pt}[0pt][0pt]{\twlrm + 3}}}
\put(286,806){\makebox(0,0)[lb]{\raisebox{0pt}[0pt][0pt]{\twlrm + 3}}}
\put(232,806){\makebox(0,0)[lb]{\raisebox{0pt}[0pt][0pt]{\twlrm (}}}
\put(167,806){\makebox(0,0)[lb]{\raisebox{0pt}[0pt][0pt]{\twlrm +}}}
\end{picture}
} \nonumber
\eqae
We will now calculate the second virial coefficient numerically and
determine the physical behaviour of the gas at low densities.

\section{The second virial coefficient}

%ffffffffffffffffffffffffffffffffffffffffffffffffffffffffffffffffffffffff
\begin{figure}[t]
\hspace{2cm}
\epsfxsize=13cm
\epsfbox{metric.potential.eps}
\caption{A plot of $F(\s)$ and $V(\s)$ for two vortices where the
inter-vortex separation is $2\s$.}
\label{fig:metpot}
\end{figure}
%fffffffffffffffffffffffffffffffffffffffffffffffffffffffffffffffffffffffff

The low density behaviour of a gas of vortices is dominated by the
second virial coefficient $B_{2} = -b_{2}$. This can be determined
from the metric and potential for two
vortices interacting in the plane, calculated numerically in
\cite{ncv}. The integrand is only non-zero in the region
where the two vortices are close, so the value for $b_{2}$
on the sphere will approach the planar value in the thermodynamic
limit. As a consequence of translational invariance and the
K\"ahler property, the metric for two vortex scattering in the
plane can be written
\eq
ds^{2} = F^{2}(\s) (d\s^{2} + \s^{2}d\theta^{2})
\eqe
where the centre of mass has been fixed at the origin to leave the vortices
at $\pm z$ with $z = \s e^{i\theta}$. $F \rightarrow 1$ exponentially
fast as $\s \rightarrow \infty$ so that the metric becomes flat as the
vortex separation increases. A plot of this metric, and the potential
for two vortices is shown in Figure \ref{fig:metpot}. When embedded
in $\reals^{3}$, this metric has the form of the familiar rounded
cone of vertex angle $\pi/2$, which was first calculated by
Samols \cite{samols}. Therefore
\eqa
B_{2} & = & -\frac{1}{2[\w]} \int g(z_{1},z_{2})dz_{1} \wedge d\zb_{1}
\wedge dz_{2} \wedge d\zb_{2} \; V_{2}(z_{1},z_{2}) \\
      & = & -\frac{1}{2[\w]} \int g(z_{1},z_{2})dz_{1} \wedge d\zb_{1}
\wedge dz_{2} \wedge d\zb_{2} \;(e^{-U_{2}(|z_{1}-z_{2}|/T} -1) \nonumber \\
      & = & -\pi \int_{0}^{\infty} d\s F^{2}(\s) \s (e^{-\frac{\l}{T}
V(\s)} -1)
\eqae
This integral can be calculated numerically and is displayed in Figure
\ref{fig:b2}. For small $\l / T$

%ffffffffffffffffffffffffffffffffffffffffffffffffffffffffffffffffffffffff
\begin{figure}
\hspace{2cm}
\epsfxsize=13cm
\epsfbox{B2.eps}
\caption{A plot of $B_{2}(\frac{\l}{T})/ \pi$. The gradient at the
origin is $\simeq 4.89$, and the solution grows exponentially for
large $-\frac{\l}{T}$.}
\label{fig:b2}
\end{figure}
%fffffffffffffffffffffffffffffffffffffffffffffffffffffffffffffffffffffffff

\eq
B_{2} = \frac{\l}{T} 4.89\pi + O(\frac{\l}{T})^{2}
\eqe
and the equation of state reduces to
\eq
\label{eq:vdwlimit}
(P - 4.89\pi \l \frac{N^{2}}{A^{2}} + O(\frac{N^{3}}{A^{3}},
\frac{\l^{2}}{T})) (A-4\pi N) = NT
\eqe
This is of the Van der Waals form $(P+b\frac{N^{2}}{A^{2}})(A-aN)=NT$
with coefficients $b=-4.89\pi\l$ and
$a=4\pi$, with the moduli space metric producing the excluded volume
effect and the static potential responsible for the deviation from an
ideal gas. It is remarkable that a complicated interaction of
extended objects in classical field theory should reduce to the
familiar Van der Waals equation, the simplest equation for a
non-ideal gas, in a low density limit. The vortices behave like
``softened'' hard discs with an extra attractive/repulsive potential.
\footnote{ ``softened'' in the sense that the vortex cores
can approach arbitrarily close if they have sufficient energy to do
so, but still exhibit an excluded volume effect.}
\par
Considering the differentiation of the pressure with respect to the
{\it area per vortex} $v=\frac{A}{N}$ we find that the compressibility
vanishes at
\eq
v = 4\pi -2B_{2}(\frac{\l}{T}) + O(v^{-1})
\eqe
Thus if a low density gas of $\l<0$ vortices had effective area less
than a certain amount close to the above value, the gas would
seek to condense (the second virial coefficient only gives an
indication of where a phase transition might be; even in a low density
gas higher order virial coefficients will become important as
we approach the transition. A phase transition is certainly expected
on physical grounds for a gas of attracting particles with no
a priori restriction on how close neighbouring particles may approach).
Plots of the equation of state for $\l<0$ and $\l>0$ are given
in Figures \ref{fig:pdplus} and \ref{fig:pdminus} respectively.

%ffffffffffffffffffffffffffffffffffffffffffffffffffffffffffffffffffffffff
\begin{figure}
\hspace{2cm}
\epsfxsize=12cm
\epsfbox{diagram.plus.eps}
\caption{A plot of pressure versus available area per vortex, $v =
A/N$ for $0.03 \le T \le 0.1$ and $\l=\frac{1}{8}$.
Recall $\r=(v-4\pi)^{-1}$. }
\label{fig:pdplus}
\end{figure}
%fffffffffffffffffffffffffffffffffffffffffffffffffffffffffffffffffffffffff

%ffffffffffffffffffffffffffffffffffffffffffffffffffffffffffffffffffffffff
\begin{figure}
\hspace{2cm}
\epsfxsize=12cm
\epsfbox{diagram.minus.eps}
\caption{A plot of pressure versus available area per vortex, $v =
A/N$ for $0.03 \le T \le 0.1$ and $\l = -\frac{1}{16}$.
The region of negative compressibility, where the gradient of the
graph is negative, is of course unphysical; condensation occurs when
the gradient vanishes.}
\label{fig:pdminus}
\end{figure}
%fffffffffffffffffffffffffffffffffffffffffffffffffffffffffffffffffffffffff

\section{The third and higher virial coefficients}

Determining the $N$th virial coefficient numerically is an expensive
business. The most obvious way to calculate it would be to determine
the metric and potential for $N$ vortices, a $2N-3$ dimensional
problem in which each point is calculated from the solution of a
second order differential equation for a static vortex configuration.
The solution of the static vortex equation is numerically difficult,
with multigrid relaxation methods giving the best convergence. The
metric and various combinations of the potential and lower order
potentials can then be integrated to give the virial coefficient.
Our only advantages are that integrand is smooth, concentrated around the
origin (where the $N$ vortices coincide), and symmetry properties may
be used to reduce the number of points needing to be calculated.
Practically, the third virial coefficient could be determined
by this ``brute force and integrate'' method, but higher virial
coefficients would certainly require more subtle methods, sacrificing
accuracy for a ``good enough'' estimate of the coefficient. For
example, Monte Carlo methods could be used to good effect.
\par
We can, however, obtain a {\it very rough} estimate of the magnitude
and sign of the higher order virial coefficients, by considering the
typical size of the integrand in the limits of $\l \ll T$ and $\l \gg T$.
We consider first small $\l/T$ when
\eq
f_{i_{1} \ldots i_{m}}(z_{i_{1}}, \ldots ,z_{i_{m}}) =
-\frac{u_{m}}{T} + O(\frac{\l}{T})^{2}
\eqe
(recall that $u_m$ is proportional to $\l$). The integrand of the virial
coefficients (\ref{eq:vcoff}) is then
\eq
X_{m} = -\frac{u_{m}}{T} + O(\frac{\l}{T})^{2}
\eqe
Now $|u_{m}|$ takes its maximum value when the vortices coincide. The
metric vanishes there on symmetry grounds, so the integrand of $B_m$ is
zero. However, the metric climbs from zero faster than the magnitude
of potential decreases (for example, see Figure \ref{fig:metpot}.) so
an estimate of the magnitude and sign of the virial coefficient may be
obtained from the value of $u_{m}$ for $m$ coincident vortices,
which we denote $u_{m}(0)$. We have determined these values for $m=2
\ldots 10$ using a one dimensional relaxation code, followed by
cubic spline integration, and tabulate the values in Table 1.
An interesting feature is that the depth
of the $m$-point potential $U_{m}$ is roughly linear with $m$, causing
$u_{m}(0)$ to change sign with $m$ and decrease in magnitude.
\par
In the case of the third virial coefficient, we see that
$\frac{u_{m}}{\l}(0)<0$, and it may also be shown that
$\frac{u_{3}}{\l}(z_{1},z_{2},z_{3})<0$ in the limit of two
vortices close, or all three well-separated. We write
\eq
B_{m} = \frac{\l}{T} \frac{1}{m(m-2)!} B_{m}^{\prime} +
O(\frac{\l}{T})^{2}
\eqe
where we estimate that $B_{m}^{\prime} = (-1)^{m} O(u_{m}(0)/\l)$.
{}From the ratio of $u_{2}(0)$ to $u_{3}(0)$, we estimate very
roughly that $-3B_{3}^{\prime} \simeq 2B_{2}^{\prime}$.
Differentiating the pressure series (\ref{eq:pressseries}) terminated
at the third virial coefficient, we find that the compressibility
vanishes at
\eq
\r_{\pm} = -\frac{B_{2}^{\prime}}{3B_{3}^{\prime}} \pm
\sqrt{\frac{B_{2}^{\prime 2}}{9B_{3}^{\prime 2}} -
\frac{T}{3\l B_{3}^{\prime}}}
\eqe
Thus for $\l >0$ vortices there is now the possibility of a
non-equilibrium phase transition at around $\r_{+}$.
For $\l<0$ we have an equilibrium phase transition between
around $\r_{+}$ and $\r_{-}$ for $|\frac{\l}{T}|
< \frac{-3B_{3}^{\prime}}{B_{2}^{\prime 2}}$, but no
phase transition if this inequality is not satisifed. However, we
should be cautious about attaching too much significance to this
result; whether these phase transitions survive the inclusion
of higher order virial coefficients depends sensitively on their
relative magnitudes; it is unlikely that there would not be a
phase transition for attracting ($\l<0$) vortices.

\begin{table}
\vspace{0.5cm}
\caption{The depths of the functions $U_{m}(0)/\l$ and $u_{m}(0)/\l$
for $m=2 \ldots 10$}
\begin{center}
\begin{tabular}{|l|r|r|}
\hline
$m$ & $U_{m}/\l$ & $u_{m}/\l$ \\
\hline
2 & 3.195 & 3.195 \\
3 & 7.357 & -2.230 \\
4 & 12.02 & 1.760 \\
5 & 16.97 & -1.491 \\
6 & 22.10 & 1.300 \\
7 & 27.31 & -1.161 \\
8 & 32.51 & 1.043 \\
9 & 37.60 & -0.902 \\
10 & 42.49 & 0.661 \\
\hline
\end{tabular}
\end{center}
\end{table}

\par
For large $\l/T$, we need to consider the cases $\l >0$ and $\l<0$
separately. For $\l <0$, $f_{12} \gg 1$ as $z_{1} \rightarrow z_{2}$
whereas $f_{123} \rightarrow -1$ as $z_{1} \rightarrow z_{2}
\rightarrow z_{3}$ (as a consequence of $u_{3}(0)/\l <0$). Thus
\eq
X_{3} \simeq -f_{12}f_{23}-f_{12}f_{13}-f_{13}f_{23}
\eqe
and hence $B_{3}$ is large and positive (in this case, $B_{2}$ is large
and negative). For $\l>0$ we find the reverse case around the centre
of the integral: $f_{12} \rightarrow -1$ and $f_{123} \gg 1$ so that
\eq
X_{3} \simeq f_{12}f_{13}f_{23}
\eqe
Therefore $B_{3} = O(1)$, negative, whereas $B_{2} = O(1)$, positive.
Analysing the pressure series we find essentially the same situation
as for small $\l/T$: evidence in favour of an equilibrium phase
transition for $\l <0$ and a non-equilibrium one for $\l>0$.

\section{Conclusions}

In this paper, we have shown that the thermodynamics of a gas of
vortices is determined by two effects: a static potential between the
vortices, producing nonzero second and higher virial coefficients, and
the curved metric on the moduli space, producing an excluded volume
effect. Any model of vortex interactions which neglects velocity
dependent terms will necessarily miss the second effect.
\par
The thermodynamics of a low density gas of vortices is clear. In the
type-II case we find no phase transition at non-zero temperature, and
this can be ultimately traced back to the exponential screening of the
repulsive static forces between the vortices. The physical properties
of the gas vary smoothly as the density is varied, and at temperatures
$T \gg \l$ the vortices behave as a Van der Waals gas. In a type-I gas
(a theoretical case when the space in which the vortices live has no
boundary) there is a phase transition between a low density gaseous state
and a condensed state where the vortices seek to superimpose
themselves into one vortex of large winding number. At temperatures
$T \gg \l$ the vortices behave as before as a Van der Waals gas.
\par
For densities of vortices approaching the Landau limit, the higher
virial coefficients have a significant effect on the properties of the
gas. As an example, we have found that on inclusion of an estimate
for the third virial coefficient there is the possibility of an
equilibrium phase transition for a type-I gas, and a non-equilibrium
phase transition at high density for a type-II gas. This also
indicates that the behaviour of a high density vortex gas will be
significantly different
to that of a low density one (as we approach the Landau limit, the
curvature of the moduli space becomes the dominant effect; the
potential is supressed, becoming constant to first order on the moduli
space. However, in this limit extra consideration should be given to
modes perpendicular to the moduli space and no detailed analysis of
these modes has yet been undertaken). The higher order virial coefficients
could be determined in a relatively straightforward way by a Monte
Carlo algorithm; we have the advantage that the integrands are smooth,
and peaked in an annular region around the origin. A better
understanding of vortex lattice ``melting'' must involve the inclusion
of higher order virial coefficients (although we note the limitations
of an analytic expansion around zero density; for example, a gas of
hard discs seems to undergo a phase transition at high densities
purely on geometrical grounds, which is not seen in the virial
coefficients \cite{harddiscs}).
\par
The theoretical behaviour of vortices in superconductors and
superfluids has been studied extensively in the field of condensed
matter. In type-II superconductors, it is observed that
the low temperature Abrikosov flux lattice can melt
as the temperature is increased, before the loss of the
superconducting state. Recent theory also suggests the
existence of intruiging extra phases, for example a transition
to a ``vortex glass'' state (in which the vortex cores do not have
sufficient energy to approach each other, and the flux tubes become
entangled) before the transition to the isotropic
state at high densities \cite{typetwo}. On general physical
grounds we would expect such a
transition to be observed in our model: in such a state, a vortex
would be effectively confined in a potential well by its nearest
neighbours, with the curvature of the moduli space enhancing the
confinement by providing a strong short range repulsion. We have
seen a indication of this behaviour from our estimates of the third
virial coefficient. However, the virial expansion is a rather blunt
tool to use, and a more detailed examination of potential and
moduli space metric is required to see if these phases, and others
such as a hexatic liquid, really exist.
\par
It is our hope that our model will provide a useful starting point
for further investigations of the dynamics of vortices; techniques for
imaging the vortices directly are improving, and we hope that vortex
dynamics and scattering may be observed in the near future
\cite{vortpics}.  Although the
calculations in this paper assumed an ideal superconductor, this is
not a limitation on the model. Terms representing disorder and vortex
pinning could be added to the potential; the utility of the
model is in its full expression of the dynamics of vortices without
using either the London limit or the lowest Landau level approximation.

\vskip 1cm
\hskip -18pt
{\bf{Acknowledgements}}
\par
I am very grateful to Nick Manton for useful discussions. This
research was supported by a grant from the Particle Physics and
Astronomy Research Council.
\eject

\end{document}